# Hopping Transport in SrTiO$_3$/Nd$_{1-x}$TiO$_3$/SrTiO$_3$ Heterostructures

Laxman Raju Thoutam[*], Jin Yue, Peng Xu and Bharat Jalan[*]

Department of Chemical Engineering and Materials Science, University of Minnesota - Twin Cities, Minneapolis, Minnesota 55455, USA

[*] Corresponding authors: lthoutam@umn.edu and bjalan@umn.edu

Electronic transport near the insulator-metal transition is investigated in the molecular beam epitaxy-grown $SrTiO_3/Nd_{1-x}TiO_3/SrTiO_3$ heterostructures using temperature dependent magnetotransport measurements. It was found that Nd-vacancies introduce localized electronic states resulting in the variable range hopping transport at low temperatures. At a fixed Nd-vacancies concentration, a crossover from Mott to Efros-Shklovskii (ES) variable range hopping transport was observed with decreasing temperature. With increasing disorder, a sign reversal of magnetoresistance from positive to negative was observed revealing interplay between intra-state interaction and the energy dependence of the localization length as a function of disorder. These findings highlight the important role of stoichiometry when exploring intrinsic effect using heterostructure and interfaces in addition to offering broad opportunity to tailor low temperature transport using non-stoichiometry defects.

Complex oxides with perovskite structure (ABO$_3$, where A and B are cations) can host large amount of disorder including dislocations, strain, and non-stoichiometry such as cation and oxygen vacancies. In particular, cation deficiency can introduce hole-type carriers and compensate for electronic carriers or can even be responsible for forming oxygen vacancies. In valence change oxides, cation vacancies can also be accommodated by valence change. It was recently shown that introduction of Nd vacancies (*x*) in Nd$_{1-x}$TiO$_3$/SrTiO$_3$ heterostructure can be accommodated by the change in Ti valence state from Ti$^{3+}$ → Ti$^{4+}$, with a direct consequence on the interfacial conductivity.[1] Cation non-stoichiometry and oxygen vacancies are also argued to play critical role in realizing 2DEGs at LaAlO$_3$/SrTiO$_3$ (LAO/STO) interfaces.[2-6] Furthermore, the phenomena like emerging magnetism, ferroelectricity in complex oxide films and heterostructures are argued to stem from defect-induced disorder.[7, 8]

Disorder in thin films can come from growth methods, epitaxial strain, interface roughness, and from the choice of growth conditions. Recent DFT study has emphasized the role of growth conditions on the defect formation energy in GdTiO$_3$ that are responsible for the p-type hopping conductivity.[9] The conductivity behavior of these oxides with different disorder level is of practical relevance as these materials are increasingly being investigated for emerging phenomena and for novel device applications such as Mott transistors.

Variable range hopping (VRH) conduction has long been studied in the context of disordered semiconductors illustrating the temperature dependence of sheet conductance at low temperatures.[10] In a compensated semiconductor, sheet conductance ($\sigma$) can follow an exponential behavior:

$$\sigma = \sigma_o \exp\left[-\left(\frac{T_0}{T}\right)^m\right] \quad (1)$$

where $\sigma_o$ is a pre-factor, $T_0$ is the characteristic temperature, and $m$ is an exponent. The value of $m$ depends on the type of transport mechanisms. For example, nearest neighbor hopping (NNH) or thermally activated transport takes a value of $m = 1$ whereas VRH yields a value of 0.25 for 3D Mott VRH, 0.33 for 2D Mott VRH, and 0.5 for Efros-Shklovskii (ES) VRH.[11] The value of $m$ can be determined using Zabrodskii's analysis, which is an unbiased and quantitative technique that linearizes the conductivity with temperature using a logarithmic derivative method by defining $w = d \ln \sigma / \ln T$.[12] The slope of $\ln w$ vs $\ln T$ plot yields $m$. The Mott VRH assumes no electron-electron interactions whereas ES VRH accounts for a long range Coulomb interaction.[10] The later results in the opening of soft Coulomb gap whose size increases with increasing disorder level (or increasing charge compensation).[10] For a given density of state, VRH can evolve from the Mott, $T^{-0.25}$, at higher temperature to the ES, $T^{-0.5}$ regime, at low temperature when Mott hopping energy becomes comparable to the ES hopping energy.[13] This crossover has been observed in the compensated semiconductors at fixed disorder level with decreasing temperature, and also by increasing disorder level at a fixed temperature.[13] With decreasing temperature, the conductivity crosses over from 3D Mott to ES with the crossover temperature given by [13]

$$T_{cross} = 16 T_{ES}^2 / T_{Mott} \quad (2)$$

In an attempt to understand role of disorder on transport in $Nd_{1-x}TiO_3/SrTiO_3$ heterostructures, we used $SrTiO_3/Nd_{1-x}TiO_3/SrTiO_3$ heterostructures with varying amount of $x$. By combining experiment and modeling of low temperature magnetoresistance (MR) measurements in the context of VRH model, we show the carrier transport is primarily governed by the ES VRH in addition to revealing an important role of intra-state correlation between localized sites on the MR behavior.

Three samples consisting of 8 unit cell (u.c.) SrTiO$_3$/ 2 u.c. Nd$_{1-x}$TiO$_3$/ 8 u.c. SrTiO$_3$ heterostructures with varying *x* were grown on LSAT (001) substrates using hybrid molecular beam epitaxy method. We have chosen this particular structure in this study because this is well characterized structurally and electronically.[1] Note the stoichiometric 8 u.c. SrTiO$_3$/ 2 u.c. Nd$_{1-x}$TiO$_3$/ 8 u.c. SrTiO$_3$ sample with *x* = 0 gives 1 e$^-$/u.c. as is expected from two polar/nonpolar interfaces.[1] Details of the growth method, structure and stoichiometry optimization are discussed elsewhere.[1, 14, 15] The value of *x* was tuned by changing Nd beam equivalent pressure (BEP) during film growth. The Ti and oxygen BEPs were kept fixed. In-situ reflection high-energy electron diffraction (RHEED) was used to ensure the crystallinity of each layer in the heterostructure prior to the growth of subsequent layer. 20 nm Al / 20 nm Ti/ 200 nm Au were sputter deposited to make electrical contacts to the sample where Al makes an ohmic contant with the sample. Temperature dependent DC transport measurements were performed in Van der Pauw geometry using Quantum Design Dynacool system. Source current were carefully chosen to ensure ohmic contacts. For MR measurements, magnetic field was swept between - 9 T and + 9 T.

Figure 1a and 1b show a schematic of the heterostructure and a representative RHEED pattern along <100>$_{substrate}$ azimuth, respectively. The appearance of the streaky pattern revealed smooth and crystalline surface morphology. Figure 1c shows temperature dependent sheet resistance (R$_s$) for 8 u.c. SrTiO$_3$/ 2 u.c. Nd$_{1-x}$TiO$_3$/ 8 u.c. SrTiO$_3$ with increasing *x* from sample A to sample C. Note that the increasing *x* leads to more disorder in the Nd$_{1-x}$TiO$_3$ layer. As shown in figure 1c, the room temperature R$_s$ of sample A and B is near the quantum resistance *h/e$^2$*. With increasing *x*, R$_s$ increases above *h/e$^2$* (sample C) at all temperatures. In particular, for T < 200 K, all samples showed an increase in R$_s$ with decreasing *T*. However, in the more disordered sample C, this resistance upturn appears even at room temperature. In the discussion below, we examine the origin of resistance upturn at low temperatures and its relationship with disorder that is created intentionally by introducing Nd-vacancies.

Figures 2a-c illustrates $\ln w$ vs. $\ln T$ plots using the Zabrodskii's analysis of sample A, B and C, respectively. Using the value of exponent $m$ in different temperature range determined from these plots, figures 2d-f shows $\ln \sigma$ vs. $T^{-m}$ plots with linear fits. We first discuss the results of sample A at a fixed $x$. Figure 2a revealed that the transport in sample A is governed by the ES VRH at T < 10 K whereas an Mott 3D VRH mechanism is operative at 10 K < T < 100 K with a crossover from ES to Mott VRH at $T_{cross} \simeq$ 10 K. The fitting of ES VRH and Mott VRH regime (figure 2d) yielded $T_{ES}$ = 15 K and $T_{Mott}$ = 311 K. Using equation 2, $T_{cross}$ can be calculated for sample A and is equal to 12 K respectively. The theoretical crossover temperature is in excellent agreement with the experiment value as marked by the blue shaded region in figure 2a. A similar crossover behavior from Mott-3D to ES VRH with decreasing temperature was seen previously in $In_xO_y$,[13] n-type CdSe,[16] $Sr_2IrO_4$,[17] and oxygen-deficient ZnO films[18] and is attributed to the formation of Coulomb gap. The characteristic temperatures, $T_{Mott}$ and $T_{ES}$ are defined as $T_{Mott} = 18/k_B N_0(E_F)\xi^3$ and $T_{ES} = 2.8e^2/\kappa\xi k_B$ respectively, where $k_B$ is Boltzmann constant, $N_0(E_F)$ is density of states at the Fermi level, $\xi$ is localization length, $e$ is electronic charge, and $\kappa$ is the dielectric constant.[11,19] The relatively low value of $T_{Mott}$ and $T_{ES}$ in sample A is reasonable given it is very close to the insulator-metal transition where $\xi$ is expected to diverge to infinity and has been observed previously.[13] The $\xi$ decreases with increasing disorder and therefore, increasing disorder is expected to increases $T_{cross}$.[10,13]

Now, we turn to the discussion of sample B and samples C with higher disorder level. Sample B yielded a similar Mott 3D VRH to ES VRH crossover and a slightly higher $T_{cross}$ = 15

K. Sample C showed the ES VRH up to T < 45 K. The characteristic temperatures, $T_{Mott}$ and $T_{ES}$, theoretical and experimental crossover temperatures obtained from the analysis of samples A, B and C are listed in Table 1. Table 1 depicts reasonably good agreement between experimental and theoretically obtained $T_{cross}$ and reveals larger $T_{cross}$ for sample B with higher disorder. A similar increase in $T_{cross}$ with increased disorder was previously seen in $In_xO_y$ films[13] and is consistent with our results. Sample C yielded much higher $T_{ES}$ = 127 K, and is consistent with the increased long range Coulomb interaction in the presence of higher disorder (or more localized sites).

To further understand the interplay between disorder and carrier localization, we performed MR measurements. Figures 3a-c show the MR of sample A, B and C respectively as a function of temperature. Sample A showed positive MR at - 9 T ≤ B ≤ + 9 T whose magnitude increases with decreasing temperature in the measured temperature range. Likewise, sample B showed a positive MR at low fields but a negative MR at higher fields. It was also found that the crossover field at which MR changes its sign decreases as temperature decreases. Sample C exhibited only negative MR at all fields. Overall, these results revealed that MR undergoes a sign reversal from positive to negative with increasing disorder from sample A to sample C. Similar trend was observed with increasing magnetic field in sample B with intermediate disorder level. These results are illustrated in figure 3d at a fixed temperature, 10 K.

We now discuss the MR results. The MR in the VRH regime can be both positive and negative. Positive MR can result from shrinkage of carrier wave function with increasing field [10, 20-22] or could be due to the intra-state correlation of carrier in the localized states.[23-25] Kurobe and Kamimura has proposed a model accounting for intra-state correlation of spins in hopping sites that could give rise to positive MR.[25] Using their numerical results, Frydman and Ovadyahu later proposed a simple analytical formula.[26] This model corroborates very well with our measurements (discussed later). According to this model, the density of states in system with intra-state correlation (in the presence of large disorder) could be thought in terms of an unoccupied state (UO), singly occupied state (SO) and doubly occupied state (DO) at the Fermi level. Consequently, four types of hopping processes can be possible viz., SO to UO, SO to SO, DO to UO and DO to SO contributing to hopping probability in the absence of any external magnetic field as schematically illustrated in figure 4a. For illustration, we only show SO → UO and SO → SO hopping processes. Applying magnetic field aligns local spins, and thus reduces the hopping probability between sites as depicted using red cross sign in figure 4b for SO → SO hopping. This behavior, in turn, results in the positive MR followed by a saturation when all spins are aligned.[25-27] This model however doesn't take into account the state-energy dependence of localization length.[25] According to this model, the positive MR or negative magnetoconductance (MC = $\frac{\Delta\sigma}{\sigma} = \frac{\sigma(H)-\sigma(0)}{\sigma(0)}$) is given by[26]

$$\frac{\Delta\sigma}{\sigma} = -A_e \frac{H^2}{H^2 + H_e^2} \quad (3)$$

$A_e$ is the saturation value, $H_e = a \frac{k_B T}{\mu_B} \left(\frac{T_{Mott}}{T}\right)^{0.5}$ is the spin alignment field, $a$ is the constant of the order unity, $k_B$ is the Boltzmann constant and $\mu_B$ is the Bohr magneton.

Negative MR in VRH can arise from orbital quantum interference effect or the state-energy dependence of $\xi$ in the presence of applied field.[28-32] The main evidence of the former comes from the presence of anisotropic MR with the field in-plane and out-of-plane of the sample.[32] Figure S1 shows the isotropic negative MR of sample C with both in-plane and out-of-plane field directions suggesting the orbital quantum interference mechanism is not operative here. As for the dependence of field on $\xi$, Fukuyama and Yoshida proposed a model explaining the negative MR in the VRH regime.[28] Schematically, we illustrate this model in figure 4b, which shows an arbitrary density of state, dependence of $\xi$ on $E_c - \varepsilon_\alpha$, where $\varepsilon_\alpha$ is state-energy of the localized electron in a SO state, the mobility edge ($E_c$) and the Fermi level ($E_F$). It is argued that an application of field can decrease electron energy by $g\mu_B H$ ($g$ is Lande factor, and H is the magnetic field) according to Zeeman effect such that $E_f = \varepsilon_\alpha - g\mu_B H$ resulting in the increase of $\xi$, and therefore, a negative MR.[25, 28] According to the Fukuyama and Yoshida model, the negative MR or the positive MC is given by[28]

$$\frac{\Delta\sigma}{\sigma} = c \left(\frac{g\mu_B H}{E_c - E_F}\right)^2 \left(\frac{T_{Mott}}{T}\right)^{0.5} \quad (4)$$

where $c$ is a constant and is equal to unity. The equations 3 and 4 explain our measurements of positive and negative MR respectively in sample A and C. Figure 4d shows the MC from the measured MR data of sample B as depicted in figure 3b. Fittings to the experimental MC data are also included as black solid lines using an equation comprising of the summation of equations 3 and 4. Our observed MC behavior is in excellent agreement quantitatively with these models. A similar positive and negative MR based on correlation effect has been previously studied both theoretically[29, 30] and experimentally in other systems including Cu-particle films,[31] heavily FeCl$_3$-doped polyacetylene,[33] and more recently, in SmNiO$_3$.[34] Future investigations will focus on detailed quantitative analysis of these MR behaviors.

In summary, electronic transport in SrTiO$_3$/Nd$_{1-x}$TiO$_3$/SrTiO$_3$ heterostructures is performed in the insulating regime near the insulator-metal transition boundary as a function of Nd-vacancies. Temperature dependent transport revealed Mott to ES VRH transition with a crossover temperature in excellent agreement with the theoretical crossover temperatures. With increasing disorder, films exhibited ES VRH transport and negative MR. The analysis of MR revealed an important interplay between intra-state correlation effect and magnetic field dependent localization length as a function of disorder and magnetic field. We argue that such insights might be useful in further study of intrinsic phenomena of oxide heterostructures as well in comparing materials grown using different synthesis approaches.

**Acknowledgements**

We thank Boris Shklovskii, and Yilikal Ayino for valuable discussions. This work was primarily funded by the U.S. Department of Energy through the University of Minnesota Center for Quantum Materials, under Grant DE-SC-0016371. J. Y. acknowledges partial support through the Young Investigator Program of the Air Force Office of Scientific Research (AFOSR) through Grant No. FA9550-16-1-0205. Parts of this work were carried out on equipment funded by the University of Minnesota NSF MRSEC.

**Figures (color online):**

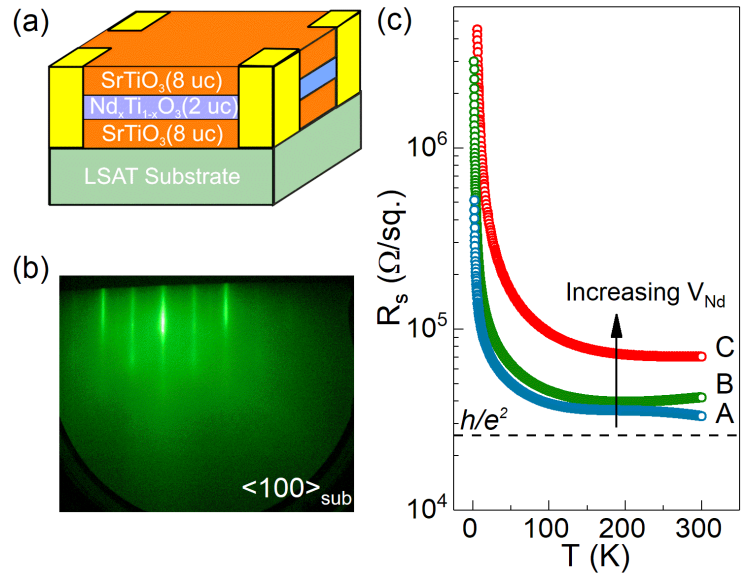

**Figure 1:** (a) Schematic of the heterostructure, (b) a representative RHEED pattern of the sample A after growth along <100> substrate azimuth. (c) Temperature dependence of $R_s$ vs. $T$ for samples A, B and C with increasing disorder level.

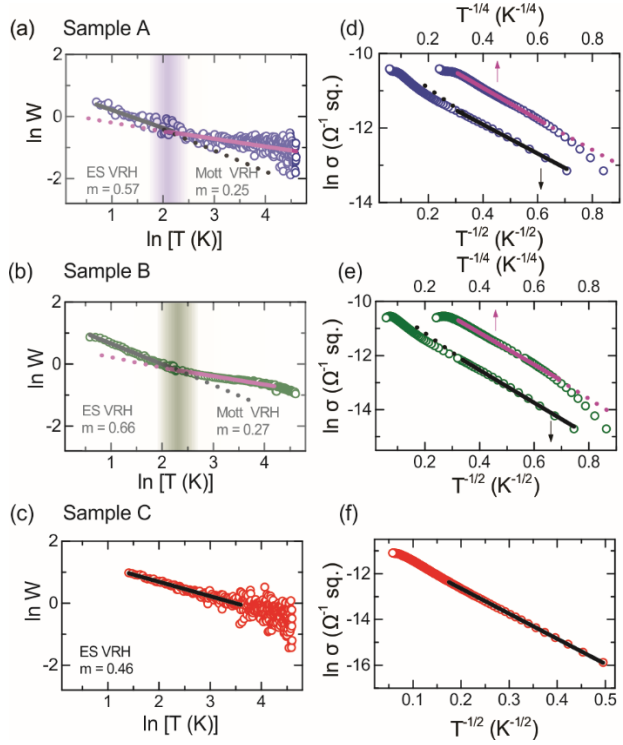

**Figure 2:** (a-c) Zabrodskii plots of sample A, B and C. The sample A and B highlights the crossover temperature between Mott VRH and ES VRH using a vertical shaded region. (d-f) ln $\sigma$ vs. $T^{-m}$ plots with linear fits using the $m$ values determined from Zabrodskii's analysis of sample A, B and C. Arrows in parts d and e indicate axes.

**Table 1:** The characteristic temperatures, $T_{Mott}$ and $T_{ES}$, theoretical and experimental crossover temperatures obtained from the analysis of samples A, B and C.

| Sample | $T_{Mott}$ (K) | $T_{ES}$ (K) | $T_{cross}$ (K) (experiment) | $T_{cross}$ (K) (theory) |
|---|---|---|---|---|
| A | 311 | 15 | ~10 | 12 |
| B | 1139 | 36 | ~15 | 18 |
| C | -- | 127 | -- | -- |

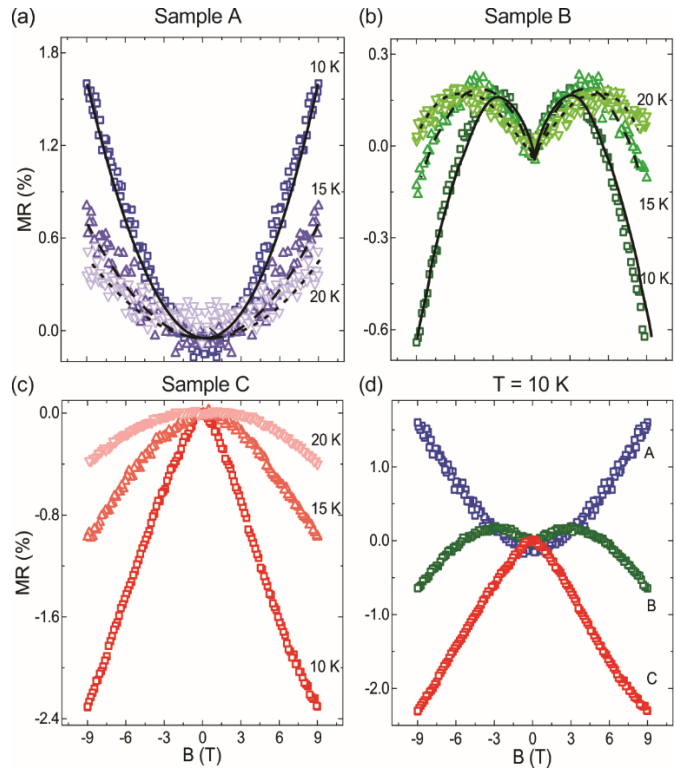

**Figure 3:** Normalized MR at 20 K, 15 K and 10 K for (a) Sample A, (b) Sample B and (c) Sample C, (d) Comparison of MR at 10K revealing a sign reversal from positive to negative values with increasing disorder from sample A to sample C. Black solid lines are guide to the eyes.

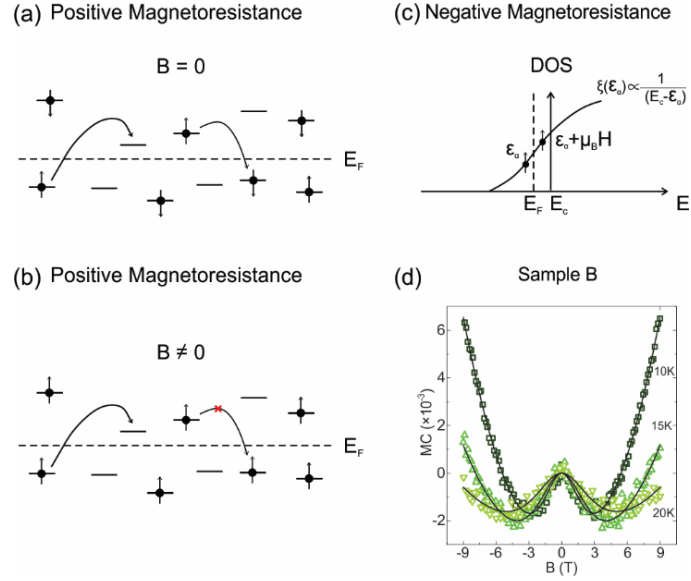

**Figure 4:** Schematic illustration of the mechanism responsible for the positive MR at (a) B = 0, and (b) B ≠ 0. Red cross sign in part b shows the transition between SO → SO localized states is energetically not possible due to spin alignment. (c) Schematic illustration of the mechanism for negative MR showing an arbitrary DOS vs. energy diagram. A formula describing the relationship between of $\xi$ and $E_c - \varepsilon_\alpha$ is included where $\varepsilon_\alpha$ is state-energy of the localized electron in a SO state, $E_c$ is the mobility edge and $E_F$ is the Fermi level. Black solid circles with an up-arrow show state-energy of a localized electron in the absence and presence of applied field, B with reference to the Fermi energy. (d) Temperature dependent magnetoconductance (MC) of sample B. Solid black lines are fits to the experimental data.